\documentclass[a4paper,12pt]{article}
\usepackage{amsfonts,amsthm,amsmath,amssymb,graphicx,hyperref,youngtab}

\def\L{\Lambda}

\newcommand{\cN}{\mathcal N}

\newcommand{\cR}{\mathcal R}

\newcommand{\cZ}{\mathcal Z}

\newcommand{\be}{\begin{equation}}
\newcommand{\bea}{\begin{eqnarray}}
\newcommand{\ee}{\end{equation}}
\newcommand{\eea}{\end{eqnarray}}

\begin{document}

\makeatletter
\@addtoreset{equation}{section}
\makeatother
\renewcommand{\theequation}{\thesection.\arabic{equation}}

\rightline{WITS-CTP-108}
\vspace{1.8truecm}

\vspace{15pt}


{\LARGE{  
\centerline{   \bf Restricted Schur Polynomials for Fermions} 
\centerline {\bf  and integrability in the su$(2|3)$ sector} 
}}  

\vskip.5cm 

\thispagestyle{empty} \centerline{
    {\large \bf Robert de Mello Koch\footnote{ {\tt robert@neo.phys.wits.ac.za}}, Pablo Diaz\footnote{ {\tt Pablo.DiazBenito@wits.ac.za}}}
   {\large \bf and Nkululeko Nokwara\footnote{\tt Nkululeko.Nokwara@students.wits.ac.za}  }}

\vspace{.4cm}
\centerline{{\it National Institute for Theoretical Physics ,}}
\centerline{{\it Department of Physics and Centre for Theoretical Physics }}
\centerline{{\it University of Witwatersrand, Wits, 2050, } }
\centerline{{\it South Africa } }

\vspace{1.4truecm}

\thispagestyle{empty}

\centerline{\bf ABSTRACT}

\vskip.4cm 

We define restricted Schur polynomials built using both fermionic and bosonic 
fields which transform in the adjoint of the gauge group $U(N)$.
We show that these operators diagonalize the free field two point function to all orders in $1/N$.
As an application of our new operators, we study the action of the one loop dilatation operator in
the su$(2|3)$ sector in a large $N$ but non-planar limit. 
The restricted Schur polynomials we study are dual to giant gravitons. 
We find that the one loop dilatation operator can be diagonalized using a double coset ansatz.
The resulting spectrum of anomalous dimensions matches the spectrum of a set of decoupled oscillators.
Finally, in an Appendix we study the action of the one loop dilatation operator in an sl$(2)$ sector.
This action is again diagonalized by a double coset ansatz.

\setcounter{page}{0}
\setcounter{tocdepth}{2}

\newpage

\tableofcontents

\setcounter{footnote}{0}

\linespread{1.1}
\parskip 4pt

{}~
{}~

\section{Introduction}

There is now convincing evidence that $\cN =4$ super-Yang-Mills theory in four 
dimensions is equivalent to ten dimensional string theory on $AdS_5 \times S^5$\cite{malda}. 
Part of specifying the equivalence entails providing a detailed identification between 
the quantum states in the  $\cN=4$ super Yang-Mills theory and objects in the dual string theory. 
Roughly, this identification is organized by the $\cR$ charge $J$ of the operators in the super Yang-Mills theory. 
For example, operators with $J\sim 1$ are identified with pointlike gravitons \cite{Gubser:1998bc,Witten:1998qj} while
operators with $J\sim\sqrt{N}$ include operators that are identified with strings\cite{Berenstein:2002jq}.
In this article we focus on certain operators with $J\sim N$ that are identified with giant gravitons\cite{mst,myers,hash}.
The study of these operators is technically interesting, since for this class of observables, the
large $N$ and the planar limits do not coincide\cite{Balasubramanian:2001nh}.

The fact that the large $N$ and the planar limits do not coincide means that summing only the planar diagrams does not produce the 
correct large $N$ value of the observable being studied - one needs to sum more than just the planar diagrams. 
This problem can be solved completely by exploiting the group theory of the symmetric and unitary groups, 
as well as the relations between them. 
Indeed, using representation theory techniques the problem of computing two point functions can be solved 
exactly in the free field limit\cite{cjr,dssi,Kimura:2007wy,BHR1,BHR2,countconst,Bhattacharyya:2008rb,Kimura:2008ac,Kimura:2009jf,Kimura:2010tx,pasram,Kimura:2012hp}. 
The next natural step is to start exploring these non-planar large $N$ limits, beyond the free theory.

Studies of the equivalence between gauge theory and gravity have contributed significantly to our understanding of ${\cal N}=4$ super Yang-Mills theory. 
Integrable structures governing the anomalous dimensions of the theory in the planar limit have been discovered\cite{mz,bks,intreview}.
This allows a complete understanding of planar two point functions and in particular, of their dependence on the coupling constant. 
Recent progress has started to provide evidence that integrability is present in other large $N$ limits of the 
theory\cite{Koch:2010gp,DeComarmond:2010ie,Carlson:2011hy,gs,Koch:2011hb,deMelloKoch:2011vn,mn,DCI}.
Given our experience with the planar limit, one might be optimistic that these new integrable sectors will again allow 
a detailed understanding, away from the free theory, of nonplanar large $N$ two point functions.
At very least, these sectors deserve further study. 
This is one of the primary motivations for this work.

Recall that operators of the super Yang-Mills theory that are dual to half-BPS giants are products of traces of powers of one matrix $Z$.
A convenient basis for these operators is provided by the Schur polynomials\cite{cjr}.
To obtain open string excitations replace some of the $Z$  matrices with another ``impurity'' matrix $Y$\cite{Balasubramanian:2002sa}.
See \cite{Aharony:2002nd,Berenstein:2003ah,Sadri:2003mx,Berenstein:2005fa,Berenstein:2006qk,dssii,bds}
for more on the construction of states corresponding to strings attached to giants.
A natural basis for these operators are the restricted Schur polynomials 
$\chi_{R,(r,s)\mu\nu}(Z,Y)$\cite{Balasubramanian:2004nb,dssi,Bhattacharyya:2008rb}.
The labels of this polynomial are $R$, a Young diagram specifying an irrep of $S_{m+n}$, a pair $(r,s)$ 
of Young diagrams for an irreducible representation (irrep) of $S_n \times S_m$ and multiplicity labels $\mu,\nu$.
The multiplicity labels resolve which $(r,s)$ irrep appears when the irrep $R$ of $S_{m+n}$ is 
decomposed under the action of the $S_n\times S_m$ subgroup. 
The action of the one-loop dilatation operator when acting on the $\chi_{R,(r,s)\mu\nu}(Z,Y)$
simplifies dramatically when lengths of distinct rows of $R$ differ by order $N$\cite{deMelloKoch:2011vn,mn}.
In this limit, the one loop dilatation operator reduces to systems of harmonic oscillators. 
In the case that $R$ has $p$ rows, the harmonic oscillator dynamics describes $p$ particles along the real line, 
with coordinates given by the lengths of the Young diagram $R$\cite{gs}.
These particles interact with quadratic two particle interaction potentials. 
The harmonic oscillator dynamics follows after diagonalization in the space of $(s,\mu,\nu)$ labels. 
This diagonalization was first considered in numerical studies of  
\cite{Koch:2010gp,DeComarmond:2010ie} for $m=2,3,4$ $Y$s.
The numerical results produce a perfect linear spectrum.
These numerical studies were then extended by an analytic approach valid
when $R$ has 2 rows or columns and $m$ is general for operators built from 2 
scalars $Z,Y$\cite{Carlson:2011hy}
and in \cite{Koch:2011jk} for operators built using 3 scalars $Z,Y,X$.
The complete action of the one loop dilatation operator for $p$ rows or columns was computed in \cite{Koch:2011hb}
and the spectrum was obtained using a numerical approach. 
The key progress was due to a new Schur-Weyl duality (this is developed further in \cite{mn}; for further
use of Schur-Weyl in AdS/CFT duality see \cite{Ramgoolam:2008yr}) which enabled both
a simple construction of the restricted Schur polynomials and the evaluation 
of the action of the one loop dilatation operator.
The numerical results for the spectrum provides a concrete realization of the Gauss Law constraints 
and motivated a very simple conjecture for the spectrum of anomalous dimensions.
Using ingredients coming from Fourier transformation on the double coset\cite{BHR1,BHR2,countconst}, 
this conjecture was proved in \cite{DCI}.  
At two loops the operators of good scaling dimension are mot modified; the anomalous dimension receives
a non-zero correction\cite{deMelloKoch:2012sv}.

The double coset ansatz and the resulting harmonic oscillator dynamics are strong signals of integrability
in this large $N$ limit. 
However, to prove that non-planar integrability is present at one loop in this sector of the theory
we need to do more. 
Apart from the scalar fields that have been considered, one needs to include the fermion 
fields and the gauge fields. 
In this article we will provide new technology that fills this important gap.
To achieve this, we explain how to construct restricted Schur polynomials which include both fermions and bosons.
These new restricted Schur polynomials continue to diagonalize the free field two point function 
(see equation (\ref{su23twopoint})) and the
number of these polynomials matches the expected number of multi field-multi trace gauge invariant operators. 
We also show how to transform between the trace basis and the basis provided by the polynomials we
construct (see equation (\ref{nicenewone})).

As a concrete application of our results we study the su$(2|3)$ sector of the theory.
The su$(2|3)$ sector consists of operators built using three complex scalars and two complex fermions. 
This sector is closed to all orders under the action of the dilatation operator. 
Useful references include \cite{beisert} where the planar dilatation operator up to three loop level was studied 
and \cite{tseytlin} who studied the coherent state action derived from the one loop dilatation operator. 
At the one loop level the dilatation operator has a rather simple action in this sector - see 
formula (2.1) of \cite{tseytlin}, or the $H_2$ piece of Table 1 of \cite{beisert}. 
In this article (section 5) we explain how to construct restricted Schur polynomials for the su$(2|3)$ 
sector and compute the action of the dilatation operator in this sector. 
We then demonstrate that the double coset ansatz of \cite{DCI} can be used to diagonalize the 
dilatation operator in this sector of the theory. 

In an appendix we show that the action of the one loop dilatation operator in an sl$(2)$ sector
is again diagonalized by a double coset ansatz.

\section{Warm up: Single Fermion}

Consider a single fermion $\psi^i_j$ transforming in the adjoint of the gauge group $U(N)$.
The relevant two point function is
\bea
 \langle \psi^i_j (\psi^\dagger)^k_l\rangle = \delta^i_l \delta^k_j
\eea
The fermionic fields are Grassman valued, so that swapping them costs a minus sign.
Our conventions for ordering the fields is as follows
\bea
  (\psi^{\otimes\, n})^I_J  = \psi^{i_1}_{j_1}\psi^{i_2}_{j_2}\cdots \psi^{i_n}_{j_n}
\eea
\bea
  (\psi^{\dagger \otimes\, n})^K_L  = \psi^{\dagger\, k_n}_{l_n}\cdots \psi^{\dagger\, k_2}_{l_2} \psi^{\dagger\, k_1}_{l_1}
  \label{daggerorder}
\eea
It is straight forward to see that
\bea
  \langle (\psi^{\otimes\, n})^I_J(\psi^{\dagger \otimes\, n})^K_L\rangle 
= \sum_{\sigma\in S_n} {\rm sgn}(\sigma)\sigma^I_L(\sigma^{-1})^K_J
\label{wickforfermions}
\eea
where ${\rm sgn }(\sigma)$ is the sign of permutation $\sigma$. 
To compute the sign of a permutation decompose it into a product of transpositions; this decomposition
is not unique. 
Then, ${\rm sgn}(\sigma)=(-1)^m$ where $m$ is the number of transpositions in the product; ${\rm sgn}(\sigma)$ is well defined,
i.e. it does not depend on the specific decomposition of $\sigma$ into transpositions.  
The ordering in (\ref{daggerorder}) is used to ensure that no extra $n$ dependent phases appear in (\ref{wickforfermions}).

The Grassman nature of $\psi$ implies that the trace of an even number of fields vanishes. For example, consider
\bea
{\rm Tr} (\psi^4)
=\psi^i_j\psi^j_k\psi^k_l\psi^l_i 
=- \psi^j_k \psi^i_j \psi^k_l\psi^l_i 
=  \psi^j_k \psi^k_l \psi^i_j \psi^l_i 
= -\psi^j_k\psi^k_l\psi^l_i\psi^i_j = - {\rm Tr} (\psi^4)
\label{odd}
\eea
Further, the product of two traces with the same number of fields in each trace vanishes. For example
\bea
  {\rm Tr} (\psi^3) {\rm Tr} (\psi^3)
= \psi^i_j\psi^j_k\psi^k_i {\rm Tr} (\psi^3)
= - \psi^i_j\psi^j_k {\rm Tr} (\psi^3) \psi^k_i
= \psi^i_j {\rm Tr} (\psi^3) \psi^j_k\psi^k_i\cr
= -{\rm Tr} (\psi^3) \psi^i_j\psi^j_k\psi^k_i
= - {\rm Tr} (\psi^3) {\rm Tr} (\psi^3)
\eea

Let us now consider polynomials built from the adjoint fermion. 
Since we want a gauge invariant operator, consider polynomials built as a linear combination of 
traces\footnote{Of course, each of these single traces in $V^{\otimes n}$ can give rise to any multitrace
structure involving the $n$ fields. Here $V$ is isomorphic to the carrier space of the fundamental representation
of $U(N)$.}
\bea
 \sum_{\sigma\in S_n}C_\sigma {\rm Tr}_{V^{\otimes n}}(\sigma \psi^{\otimes \, n})
\eea
By changing summation variables to $\gamma^{-1}\sigma\gamma$ and
using the Grassman nature of the fermionic fields we find
\bea
 \sum_{\sigma\in S_n}C_\sigma {\rm Tr}_{V^{\otimes n}}(\sigma \psi^{\otimes \, n})
&=& \sum_{\sigma\in S_n}C_{\gamma^{-1}\sigma\gamma} {\rm Tr}_{V^{\otimes n}}(\sigma \gamma \psi^{\otimes \, n}\gamma^{-1})\cr
&=& \sum_{\sigma\in S_n}C_{\gamma^{-1}\sigma\gamma} {\rm sgn}(\gamma){\rm Tr}_{V^{\otimes n}}(\sigma \psi^{\otimes \, n} )
\eea
Thus, the coefficients used to define our polynomial must obey
\bea
  C_{\gamma^{-1}\sigma\gamma}={\rm sgn}(\gamma) C_\sigma
  \label{keyproperty}
\eea
A natural way to achieve this is to consider
\bea
\chi^F_R(\psi) = \sum_{\alpha\in S_n} S^{[1^n]\,R\, R}_{\quad\,\, m'\, m} \Gamma^R_{m\, m'} ( \alpha ) {\rm Tr}_{V^{\otimes n}} (\alpha \psi^{\otimes n})
\label{NewOperators}
\eea
where $\Gamma^R_{m\,m'}(\alpha )$ is the matrix representing $\alpha\in S_n$ in irrep $R$ and
$S^{[1^n]\,R\, R}_{\quad\,\, m\, m'}$ is the Clebsch-Gordan coefficient for $R \times R$ to couple to
the antisymmetric irrep $[1^n ]$. This formula can be viewed, as a ``degeneration'' of the operators
constructed in \cite{BHR1,BHR2}
\bea
  \sum_{\sigma\in S_n} B_{j\beta}S^{\tau ;\L \,\, R R }_{\,\,j\quad p\,\, q}  \Gamma^{ \Lambda }_{ p q } (\sigma)
  {\rm Tr}_{V^{\otimes n}}(\sigma {\bf X}^{\mu\,\,\otimes \, n} )
  \label{BHRops}
\eea
which provides a basis for $M$ species of complex matrix (different species indexed by $\mu$). The basis thus obtained
has good $U(M)$ quantum numbers (see the first formula in section 1.1 of \cite{BHR1}). 
Since $[1^n]$ appears only once in $R \otimes R$ the analog of the multiplicity label $\tau$ which appears
in (\ref{BHRops}) is not needed in (\ref{NewOperators}).
(\ref{NewOperators}) is the simplest way to turn the ``counting formula'' (eqn. 106 of \cite{BHR1})
into a ``construction formula''.

To simplify the notation write the Schur polynomials for fermions as
\bea
\chi^F_R(\psi) = \sum_{\sigma\in S_n}C_\sigma {\rm Tr}_{V^{\otimes n}}(\sigma \psi^{\otimes \, n})
               = \sum_{\alpha\in S_n} {\rm Tr}\left(O\Gamma^R(\alpha)\right){\rm Tr}_{V^{\otimes n}}(\alpha\psi^{\otimes n})
\label{MoreNewOperators}
\eea
where $O_{m\, m'}=S^{[1^n]\,R\, R}_{\quad\,\, m\, m'}$. The Clebsch-Gordan coefficients of the symmetric group 
obey (see formula 7-186 of \cite{Hammermesh})
\bea
  \Gamma^{\mu}_{ij}(\sigma)\Gamma^{\nu}_{kl}(\sigma)S^{\lambda\tau_\lambda\,\mu\, \nu}_{s\,\quad j\, l}
    =\Gamma^{\lambda\tau_\lambda}_{s's}(\sigma)S^{\lambda\tau_\lambda\,\mu\, \nu}_{s'\,\quad i\, k}
\eea
Lets specialize this to our problem. Replace $\mu,\nu$ by $R$ and $\lambda$ by $[1^n]$. There is no need for the
multiplicity label $\tau_\lambda$. Also, because $[1^n]$ is 1 dimensional there is no need for indices $s,s'$ and
we replace $\Gamma^{\lambda\tau_\lambda}_{s's}(\sigma)\to {\rm sgn}(\sigma)$. 
The equation for the Clebsch-Gordan coefficients becomes
\bea
  \Gamma^{R}_{ij}(\sigma)\Gamma^{R}_{kl}(\sigma)S^{[1^n]\,R\, R}_{\quad\,\, j\, l}={\rm sgn}(\sigma) S^{[1^n]\,R\, R}_{\quad\,\, i\, k}
\eea
which, since we may assume without loss of generality that we have an orthogonal representation, implies that
\bea
  S^{[1^n]\,R\, R}_{\quad\,\, m\, l}\Gamma^{R}_{lk}(\sigma) ={\rm sgn}(\sigma)\Gamma^{R}_{mi}(\sigma)S^{[1^n]\,R\, R}_{\quad\,\, i\, k}
\eea
This proves that
\bea
  \Gamma^S(\sigma )O={\rm sgn}(\sigma )\, O\Gamma^S(\sigma )
\label{anticom}
\eea
Clearly then, $O^2$ commutes with every element of the group and is, by Schur's Lemma, proportional to the identity matrix.
Thus, (perhaps after a normalization) we have
\bea
  O^2 ={\bf 1}
\eea
This immediately implies that characters for all odd elements (those with sign -1) of the symmetric group vanish since
\bea
  {\rm Tr}(\Gamma^R (\sigma ))&=&{\rm Tr}(O^2 \Gamma^R (\sigma ))={\rm sgn}(\sigma){\rm Tr}(O \Gamma^R (\sigma ) O)\cr
                &=&{\rm sgn}(\sigma){\rm Tr}(OO \Gamma^R (\sigma ))={\rm sgn}(\sigma){\rm Tr}(\Gamma^R (\sigma ))
\eea
where we used (\ref{anticom}) and then cyclicity of the trace.
The representation $s^T$ which is conjugate to $s$ is defined by flipping the Young diagram as shown
\bea
s=\yng(4,2,2,1)\qquad s^T=\yng(4,3,1,1)
\eea
$O$ can only be non-zero for self conjugate irreps because it is only for these that 
the characters of all odd elements vanish. 
Indeed $S^{[1^n]\,R\, R}_{\quad\,\, m\, m'}$ is only non-zero for self conjugate irreps.

Of course, the above observations all follow from
\bea
  C_{\gamma^{-1}\sigma\gamma}&=&{\rm Tr}\left(O\Gamma^R(\gamma^{-1}\sigma\gamma)\right)\cr
&=&{\rm Tr}\left(O\Gamma^R(\gamma^{-1})\Gamma^R(\sigma)\Gamma^R(\gamma)\right)\cr
&=&{\rm sgn}(\gamma){\rm Tr}\left(\Gamma^R(\gamma^{-1})O\Gamma^R(\sigma)\Gamma^R(\gamma)\right)\cr
&=&{\rm sgn}(\gamma) C_\sigma
\eea
which proves that the coefficients of our polynomials do indeed obey (\ref{keyproperty}).

Spelling out index structures, our conventions are
\bea
  \chi_R(\psi)={1\over n!}\sum_{\sigma\in S_n} {\rm Tr}(O\Gamma^R(\sigma))\psi^{i_1}_{i_{\sigma (1)}}\cdots\psi^{i_n}_{i_{\sigma (n)}}
\eea
\bea
  \chi_R^\dagger (\psi)={1\over n!}\sum_{\sigma\in S_n} {\rm Tr}(\Gamma^R(\sigma)O)\psi^{\dagger\,\,i_n}_{i_{\sigma (n)}}
  \cdots\psi^{\dagger\,\,i_1}_{i_{\sigma (1)}}
\eea
A difference between working with fermionic variables as opposed to bosonic variables, is that fermionic variables anticommute.
Thus different choices of how we populate the slots with fermionic fields can differ by a sign.
It is for this reason that we must spell things out.
We will now compute the two point function
\bea
  \langle \chi_R \chi_S^\dagger\rangle &=& {1\over (n!)^2}\sum_{\sigma,\rho,\gamma\in S_n} {\rm Tr}(O\Gamma^R(\sigma))
  {\rm Tr}(\Gamma^S(\rho)O){\rm sgn}(\gamma ){\rm Tr}_{V^{\otimes n}}(\gamma\sigma\gamma^{-1}\rho)\cr
  &=& {1\over (n!)^2}\sum_{\beta,\rho,\gamma\in S_n} {\rm Tr}(O\Gamma^R(\gamma^{-1}\beta\gamma))
  {\rm Tr}(\Gamma^S(\rho)O){\rm sgn}(\gamma ){\rm Tr}_{V^{\otimes n}}(\beta\rho)\cr
  &=& {1\over (n!)^2}\sum_{\beta,\rho,\gamma\in S_n} {\rm Tr}(O\Gamma^R(\beta))
  {\rm Tr}(\Gamma^S(\rho)O){\rm Tr}_{V^{\otimes n}}(\beta\rho)\cr
  &=& {1\over n!}\sum_{\psi,\rho\in S_n} {\rm Tr}(O\Gamma^R(\psi)\Gamma^R(\rho^{-1}))
  {\rm Tr}(\Gamma^S(\rho)O){\rm Tr}_{V^{\otimes n}}(\psi)\cr
  &=& {\delta_{RS}\over d_R}\sum_{\psi\in S_n} {\rm Tr}(\Gamma^R(\psi)) {\rm Tr}_{V^{\otimes n}}(\psi)\cr
  &=& \delta_{RS}f_R
\eea

This completes the construction of Schur polynomials for a single fermion. 
We now want to construct restricted Schur polynomials for an abitrary number of fermionic and bosonic matrix flavors. 
We will first consider the counting of these operators. 
For the counting relevant for a single fermionic variable see eqn. 106 of \cite{BHR1}. As we commented above, our
construction formula is motivated by this counting and the number of operators we have matches this counting.

\section{Counting}

We will start with a quick review of counting for bosons in the next subsection\cite{storm}. After this warm up we consider the
counting of operators built from fermions and bosons.

\subsection{Warm up: bosons}

We will count the number of operators built with $k$ species of bosonic fields. This should equal the number of
restricted Schur polynomials $\chi_{R,(r_1,r_2,\cdots,r_k)}$.

Start from the $U(N)$ partition function as quoted in \cite{Dolan}, in formula (3.7), for the case of $k$ bosonic fields
\bea
\cZ_{U(N)}(t)={1\over (2\pi i)^N N!}\oint \prod_{i=1}^N \, {dz_i\over z_i} \,
               \Delta (z)\Delta (z^{-1})\prod_{j=1}^k\prod_{r,s=1}^N {1\over 1-t_j z_r z^{-1}_s}
\label{BosonZ}
\eea
Use the Cauchy-Littlewood formula
\bea
\prod_{i=1}^L \prod_{j=1}^M {1\over 1-x_i y_j}=\sum_{r\, , \,\, l(r) \le {\rm min}(L,M)}\chi_{r}(x)\chi_{r}(y)
\label{CauchyLittlewood}
\eea
to rewrite (\ref{BosonZ}) as
\bea
\cZ_{U(N)}(t)={1\over (2\pi i)^N N!}\oint \, \prod_{i=1}^N \, {dz_i\over z_i} \,
               \Delta (z)\Delta (z^{-1})\prod_{j=1}^k \,\,
               \sum_{r_j\, , \,\, l(r_j) \le N}\chi_{r_j}(t_j z)\chi_{r_j}(z^{-1})
\eea
Now, since the Schur polynomial $\chi_r (z)$ is a homogeneous polynomial of order $|r|\, \equiv\,$ the number of boxes in $r$,
we know that
\bea
\cZ_{U(N)}(t)={1\over (2\pi i)^N N!}\oint\,\prod_{i=1}^N \, {dz_i\over z_i}
               \Delta (z)\Delta (z^{-1})\prod_{j=1}^k \,\,
               \sum_{r_j\, , \,\, l(r_j) \le N} (t_j)^{|r_j|} \chi_{r_j}(z)\chi_{r_j}(z^{-1})
\eea
Using the Littlewood-Richardson rule to perform the product of the Schur polynomials we find
\bea
\cZ_{U(N)}(t)=&&{1\over (2\pi i)^N N!}\sum_{r_1,...,r_{k+2},\, l(r_i)\le N}
              (t_1)^{|r_1|}(t_2)^{|r_2|}\cdots (t_k)^{|r_k|}
              g(r_1,r_2,\cdots,r_k,r_{k+1})\cr
              &&\times g(r_1,r_2,\cdots,r_k,r_{k+2}) 
              \oint \, \prod_{i=1}^N \, {dz_i\over z_i} \, \Delta (z)\Delta (z^{-1}) \chi_{r_{k+1}}(z)\chi_{r_{k+2}}(z^{-1})
\eea
Now,
\bea
  \langle g,h\rangle_N\equiv {1\over (2\pi i)^N N!} \oint \, \prod_{i=1}^N \, {dz_i\over z_i} \, \Delta (z)\Delta (z^{-1}) g(z)h(z^{-1})
\eea
and $\langle\chi_r ,\chi_t\rangle_N =\delta_{rt}$ so that
\bea
\cZ_{U(N)}(t)=\sum_{r_1,...,r_{k},R,\,\, l(r_i)\le N, \,\, l(R)\le N}
              (t_1)^{|r_1|}(t_2)^{|r_2|}\cdots (t_k)^{|r_k|}
              (g(r_1,r_2,\cdots,r_k,R))^2
\label{bosoniccount}
\eea
From the coefficient of $(t_1)^{n_1}(t_2)^{n_2}\cdots (t_k)^{n_k}$ we learn how many operators can be built using $n_k$ fields
of species $k$. This is in turn equal to the number of restricted Schur polynomials $\chi_{R,(r_1,r_2,\cdots,r_k)}$ with
$r_i\vdash n_i$ and $R\vdash n_1+n_2+\cdots+n_k$\cite{storm}.

\subsection{One fermion, one boson}

We will count the number of operators built with one bosonic species and one fermionic species of field. 
Use $r$ for the bosonic Young diagram and $s$ for the fermionic Young diagram.

Start from the $U(N)$ partition function as quoted in \cite{Dolan}, in formula (3.13), for the case of one bosonic field
and one fermionic field
\bea
\cZ_{U(N)}(f,b)={1\over (2\pi i)^N N!}\oint\,\prod_{i=1}^N \, {dz_i\over z_i} \,
               \Delta (z)\Delta (z^{-1})\prod_{r,s=1}^N {1-f z_r z_s^{-1} \over 1-b z_r z^{-1}_s}
\label{BosonFermion}
\eea
Use the Cauchy-Littlewood formula (\ref{CauchyLittlewood}) and Littlewood's formula
\bea
\prod_{i=1}^L \prod_{j=1}^M (1+x_i y_j)=\sum_{s\, , \,\, l(s) \le L \,\, l(s^T) \le M}\chi_{s}(x)\chi_{s^T}(y)
\label{Littlewood}
\eea
where $s^T$ is conjuagte to $s$, to rewrite (\ref{BosonFermion}) as
\bea
\cZ_{U(N)}(f,b)&=&{1\over (2\pi i)^N N!}\oint\, \prod_{i=1}^N \, {dz_i\over z_i} \,
               \Delta (z)\Delta (z^{-1})\cr
               &&\times\sum_{r,\,s , \,\, l(r) \le N\,\, l(s) \le N \,\, l(s^T) \le N}
               \chi_{r}(b z)\chi_{r}(z^{-1})
               \chi_{s}(f z)\chi_{s^T}(z^{-1})
\eea
Now, since the Schur polynomial $\chi_t (z)$ is a homogeneous polynomial of order $|t|\, \equiv\,$ the number of boxes in $t$,
we know that
\bea
\cZ_{U(N)}(f,b)&=&{1\over (2\pi i)^N N!}\oint \,\prod_{i=1}^N \, {dz_i\over z_i} \,
               \Delta (z)\Delta (z^{-1})\cr
             &&\times\sum_{r,\,s , \,\, l(r) \le N\,\, l(s) \le N \,\, l(s^T) \le N} b^{|r|} f^{|s|}
               \chi_{r}(z)\chi_{r}(z^{-1})
               \chi_{s}(z)\chi_{s^T}(z^{-1})
\eea
Using the Littlewood-Richardson rule to perform the product of the Schur polynomials we find
\bea
\cZ_{U(N)}(f,b)&=&{1\over (2\pi i)^N N!}\oint \,\prod_{i=1}^N \, {dz_i\over z_i}
               \Delta (z)\Delta (z^{-1}) \sum_{r,\,s , \,\, l(r) \le N\,\, l(s) \le N\,\, l(s^T) \le N}\cr
               &&\times  \sum_{R_1,\, R_2\,\, l(R_i)\le N}b^{|r|} f^{|s|}
               g(r,s,R_1)g(r,s^T,R_2)\chi_{R_1}(z)\chi_{R_2}(z^{-1})
\eea
Now, again using $\langle\chi_r ,\chi_t\rangle_N =\delta_{rt}$ we have
\bea
\cZ_{U(N)}(f,b)= \sum_{r,\,s , \,\, l(r) \le N\,\, l(s) \le N\,\, l(s^T) \le N}\sum_{R\,\, l(R)\le N}
              b^{|r|}f^{|s|} g(r,s,R)g(r,s^T,R)
\label{FrmnBsn}
\eea

The fermionic statistics are reflected in this answer. Since the fermionic matrix is a matrix of Grassman variables any product
with more than $N^2$ factors of the fermionic matrix will vanish. Note that since both $l(s)\le N$ and $l(s^T)\le N$, $s$
can have at most $N^2$ boxes, i.e. we never get operators with a product of more than $N^2$ factors of the fermionic matrix. 
Note also that, in general
\bea
  g(r,s,R)  \ne  g(r,s^T,R)
\eea
so that this counting is genuinely different to (\ref{bosoniccount}).

\subsection{Fermions and bosons}

We will now count the number of operators built with $n_b$ species of bosonic fields and $n_f$ species of fermionic fields.

Start from the $U(N)$ partition function as quoted in \cite{Dolan}, in formula (3.13), for the case of $n_b$ bosonic fields
and $n_f$ fermionic fields
\bea
\cZ_{U(N)}(f,b)={1\over (2\pi i)^N N!}\oint\,\prod_{i=1}^N \, {dz_i\over z_i} \,
               \Delta (z)\Delta (z^{-1})\prod_{j=1}^{n_f}\prod_{k=1}^{n_b}\prod_{r,s=1}^N {1-f_j z_r z_s^{-1} \over 1-b_k z_r z^{-1}_s}
\label{BosonZFermionZ}
\eea
Using the Cauchy-Littlewood formula (\ref{CauchyLittlewood}) and Littlewood's formula (\ref{Littlewood})
we can rewrite (\ref{BosonZFermionZ}) as
\bea
\cZ_{U(N)}(f,b)&=&{1\over (2\pi i)^N N!}\oint\, \prod_{i=1}^N \, {dz_i\over z_i} \,
               \Delta (z)\Delta (z^{-1})\prod_{j=1}^{n_f}\prod_{k=1}^{n_b}
               \sum_{r_k,s_j \, , \,\, l(r_k) \le N\,\, l(s_j) \le N \,\, l(s_j^T) \le N}\cr
               &&\times \chi_{r_k}(b_k z)\chi_{r_k}(z^{-1})\chi_{s_j}(f_j z)\chi_{s^T_j}(z^{-1})
\eea
Now, again, since the Schur polynomial $\chi_t (z)$ is a homogeneous polynomial of order $|t|$ and 
using the Littlewood-Richardson rule to perform the product of the Schur polynomials we find
\bea
\cZ_{U(N)}(f,b)&=&{1\over (2\pi i)^N N!}\sum_{r_1,\cdots,r_{n_b}\,\,l(r_a)\le N}\,\,
                                        \sum_{s_1,\cdots,s_{n_f}\,\,l(s_b)\le N\,\,l(s^T_b)\le N}\,\,
                                        \sum_{R_1,R_2\,\, l(R_i)\le N}\cr
                 &&\times(f_1)^{|s_1|}\cdots (f_{n_f})^{|s_{n_f}|}(b_1)^{|r_1|}\cdots (b_{n_b})^{|r_{n_b}|}\cr
                 &&\times g(r_1,\cdots,r_{n_b},s_1,\cdots,s_{n_f},R_1) 
                          g(r_1,\cdots,r_{n_b},s_1^T,\cdots,s^T_{n_f},R_2)\cr
                 && \times \oint \,\prod_{i=1}^N \, {dz_i\over z_i}
               \Delta (z)\Delta (z^{-1})\chi_{R_1}(z)\chi_{R_2}(z^{-1})
\eea
Now, again using $\langle\chi_t ,\chi_u\rangle_N =\delta_{tu}$ we have
\bea
\cZ_{U(N)}(f,b)&=& \sum_{r_1,\cdots,r_{n_b}\,\,l(r_a)\le N}\,\,
                 \sum_{s_1,\cdots,s_{n_f}\,\,l(s_b)\le N\,\,l(s^T_b)\le N}\,\,\sum_{R \,\, l(R)\le N}
                 (f_1)^{|s_1|}\cdots (f_{n_f})^{|s_{n_f}|}(b_1)^{|r_1|}\cdots (b_{n_b})^{|r_{n_b}|}\cr
                && \times g(r_1,\cdots,r_{n_b},s_1,\cdots,s_{n_f},R)
                          g(r_1,\cdots,r_{n_b},s_1^T,\cdots,s^T_{n_f},R)
\eea
Note that again, in general
\bea
  g(r_1,\cdots,r_{n_b},s_1,\cdots,s_{n_f},R) \ne g(r_1,\cdots,r_{n_b},s_1^T,\cdots,s^T_{n_f},R)
\eea

\section{Restricted Schurs for su$(2|3)$}

Having learnt how to count the operators built using both fermionic and bosonic fields, we now consider their
construction. 

\subsection{Preliminary Comments}

How many times does $\big[1^n\big]$ appear in $s\otimes s^T$? In general we have
\bea
  s\otimes s^T =\oplus_t a_t\, t
\eea
To determine the positive integer $a_t$ with $t=[ 1^n ]$ start from the formula for the character of
a direct product representation
\bea
  \chi_s(g)\chi_{s^T}(g)=\sum_t\, a_t\, \chi_t (g)
\eea
and use the character orthogonality relation
\bea
  {1\over |{\cal G}|}\sum_{g\in {\cal G}}\chi_R(g)\chi_S(g^{-1})=\delta_{RS}
\eea
to obtain
\bea
  a_{[ 1^n ]}&=&{1\over |{\cal G}|}\sum_g\chi_s (g)\chi_{s^T}(g)\chi_{[1^n]}(g^{-1})\cr
&=&{1\over |{\cal G}|}\sum_g\chi_s (g)\chi_{s^T}(g){\rm sgn}(g)\cr
&=&{1\over |{\cal G}|}\sum_g\chi_s (g)\chi_{s}(g)\cr
&=&{1\over |{\cal G}|}\sum_g\chi_s (g^{-1})\chi_{s}(g)\cr
&=&1
\eea
Thus, there is no need for a multiplicity label. 
In the above we used the fact that $\chi_{s^T}(g){\rm sgn}(g)=\chi_{s}(g)$ and $\chi_{s}(g)=\chi_{s}(g^{-1})$.
In this case Hammermesh's formula reads
\bea
\Gamma^{s}_{ij}(\sigma)\Gamma^{s^T}_{kl}(\sigma)S^{[1^n]\,\,s\, s^T}_{\,\qquad j\, l} 
={\rm sgn}(\sigma)S^{[1^n]\, \,s\, s^T}_{\,\qquad i\, k}
\eea
Using the fact that we have an orthogonal rep we find
\bea
  \Gamma^{s}_{ij}(\sigma)\hat{O}_{jp}= {\rm sgn} (\sigma)\hat{O}_{ik}\Gamma^{s^T}_{kp}(\sigma)
\eea
where
\bea
  \hat{O}_{jl}=S^{[1^n]\, s\, s^T}_{\,\qquad j\, l}
\eea
$\hat{O}_{jl}$ is a map from $s^T$ to $s$. 
$\hat{O}^T\hat{O}$ maps from $s^T$ to $s^T$ and it commutes with all elements
of the group. Thus it is proportional to the identity.
$\hat{O}\hat{O}^T$ maps from $s$ to $s$ and it commutes with all elements
of the group. Thus it is also proportional to the identity.
By normalizing correctly we can choose
\bea
   \hat{O}^T\hat{O}={\bf 1}_{s^T}\qquad \hat{O}\hat{O}^T={\bf 1}_{s}
\eea
In what follows we will subduce two irreps from $R$, namely $(r,s\alpha)$ and $(r,s^T\beta)$. 
$\alpha$ and $\beta$ are multiplicity labels. 
The way we build the operators (\ref{symmetricgroupoperators})
that are used in the restricted Schur polynomials is to pull boxes off $R$ leaving $r$ behind. 
We then assemble the removed boxes to obtain $s$ or $s^T$, with some multiplicity label. 
To reflect the fact that the multiplicity ``belongs to'' $s$
and $s^T$ (and not to $[1^n]$) in our notation, we will denote
\bea
  \hat{O}_{jl}(s\alpha;s^T\beta)\equiv S^{[ 1^n]\, s,\alpha\,\,\,\, s^T,\beta}_{\,\qquad j\quad\, l}
  \label{generalO}
\eea
Making use of the operators (\ref{generalO}) is the simplest way to turn the counting formula (\ref{FrmnBsn})
into a construction formula.

\subsection{Construction}

In terms of the operators
\bea
  P_{R,(r,s)\alpha\beta}={\bf 1}_r\otimes \hat{O}(s\alpha;s^T\beta)\qquad
  P^\dagger_{R,(r,s)\alpha\beta}={\bf 1}_r\otimes \hat{O}(s^T\beta ; s\alpha)
  \label{symmetricgroupoperators}
\eea
we can write the restricted Schur polynomials as
\bea
   \chi_{R,(r,s)\alpha\beta}(Z,\psi)={1\over n!m!}\sum_{\sigma\in S_{n+m}}{\rm Tr}(P_{R,(r,s)\alpha\beta}\Gamma^R(\sigma))
                \psi^{i_1}_{i_{\sigma(1)}} \cdots \psi^{i_m}_{i_{\sigma(m)}} 
                Z^{i_{m+1}}_{\sigma (m+1)} \cdots Z^{i_{m+n}}_{\sigma (m+n)}\cr
   \chi^\dagger_{R,(r,s)\alpha\beta}(Z,\psi)={1\over n!m!}\sum_{\sigma\in S_{n+m}}{\rm Tr}(P^\dagger_{R,(r,s)\alpha\beta}\Gamma^R(\sigma))
                \psi^{\dagger i_m}_{i_{\sigma(m)}} \cdots \psi^{\dagger i_1}_{i_{\sigma(1)}} 
                Z^{i_{m+1}}_{\sigma (m+1)} \cdots Z^{i_{m+n}}_{\sigma (m+n)}\cr
\eea
The specific choice of which slots we use for $Z$ or $\psi$ is unimportant - they are related by
performing an inner automorphism on $S_{n+m}$ - which is a symmetry of the Schur polynomial.
The ordering of the $Z$ fields is completely arbitrary. The ordering of the $\psi$ fields fixes
a sign. Note that
\bea
  P_{R,(r,s)\alpha\beta}\Gamma^r(\sigma_1)\circ \Gamma^{s^T}(\sigma_2) 
= {\rm sgn}(\sigma_2)\Gamma^r(\sigma_1)\circ \Gamma^{s}(\sigma_2)P_{R,(r,s)\alpha\beta}
\eea
This imples that $P_{R,(r,s)\alpha\beta}$ is an intertwining map in the carrier space of $R$
from the subspace $(r,s^T)$ to the subspace $(r,s)$. 
Further
\bea
  P_{R,(r,s)\alpha\beta}P_{T,(t,u)\delta\gamma}^\dagger = \delta_{RT}\delta_{rt}\delta_{su}\delta_{\beta\gamma}\bar{P}_{R,(r,s)\alpha\delta}
\eea
where
\bea
   \bar{P}_{R,(r,s)\alpha\gamma}={\bf 1}_r\otimes \sum_j |s,\alpha ; j\rangle\langle s,\gamma ; j|
\eea
It is now straight forward to show that
\bea
\langle \chi_{R_1,(r_1,s_1)\alpha\beta}(Z,\psi)\chi^\dagger_{R_2,(r_2,s_2)\gamma\delta}(Z,\psi)\rangle
= \delta_{R_1 R_2}\delta_{r_1 r_2}\delta_{s_1 s_2}\delta_{\beta\delta}\delta_{\alpha\gamma}
{f_{R_1} {\rm hooks}_{R_1}\over {\rm hooks}_{r_1}{\rm hooks}_{s_1}}
\eea

The generalization to many fermions and bosons is straight forward. For the su$(2|3)$ sector in particular
we have
\bea
  P_{R,(\vec{r},\vec{s})\vec{\alpha}\vec{\beta}}&=&{\bf 1}_{r_1}\otimes 
                         \sum_j |r_2,\alpha_1 ; j\rangle\langle r_2,\beta_1 ; j|\otimes
                         \sum_k |r_3,\alpha_2 ; k\rangle\langle r_3,\beta_2 ; k|\otimes\cr
                    & &     \hat{O}(s_1\alpha_3 ;s_1^T\beta_3)\otimes
                         \hat{O}(s_2\alpha_4;s_2^T\beta_4)
\label{completeprojector}
\eea
We have written this with a specific procedure for the construction of $P_{R,(\vec{r},\vec{s})\vec{\alpha}\vec{\beta}}$ in mind.
We imagine that boxes are removed from $R$ until $r_1$ is obtained. The boxes removed are then assembled
to produce the representations $r_2,r_3,s_1,s_2$. Following this construction, $r_1$ has no multiplicity, $r_2$ has multiplicities
$\alpha_1$ and $\beta_1$, $r_3$ has multiplicity $\alpha_2$ and $\beta_2$, $s_1$ has multiplicity $\alpha_3$, $s_1^T$ has 
multiplicity $\beta_3$, $s_2$ has multiplicity $\alpha_4$ and $s_2^T$ has multiplicity $\beta_4$. Our conventions
for the ordering of the fermionic fields are
\bea
   \chi_{R,(\vec{r},\vec{s})\vec{\alpha}\vec{\beta}}(Z,X,Y,\psi_1,\psi_2)
        ={1\over n_1!n_2!n_3!m_1!m_2!}\sum_{\sigma\in S_{n_1+n_2+n_3+m_1+m_2}}{\rm Tr}(P_{R,(\vec{r},\vec{s})\vec{\alpha}\vec{\beta}}\Gamma^R(\sigma))\times\cr
                \times\psi^{i_1}_{1\, i_{\sigma(1)}} \cdots \psi^{i_{m_1}}_{1\, i_{\sigma(m_1)}}
                \psi^{i_{m_1+1}}_{2\, i_{\sigma(m_1+1)}} \cdots \psi^{i_{m_1+m_2}}_{2\, i_{\sigma(m_1+m_2)}} 
                X^{i_{m_1+m_2+1}}_{\sigma (m_1+m_2+1)} \cdots \cr
   \chi^\dagger_{R,(\vec{r},\vec{s})\vec{\alpha}\vec{\beta}}(Z,X,Y,\psi_1\psi_2)
        ={1\over n_1!n_2!n_3!m_1!m_2!}\sum_{\sigma\in S_{n_1+n_2+n_3+m_1+m_2}}
         {\rm Tr}(P^\dagger_{R,(\vec{r},\vec{s})\vec{\alpha}\vec{\beta}}\Gamma^R(\sigma))\cr
                \psi^{\dagger i_{m_1+m_2}}_{2\,i_{\sigma(m_1+m_2)}} \cdots
                \psi^{\dagger i_{1+m_1}}_{2\,i_{\sigma(1+m_1)}}\psi^{\dagger i_{m_1}}_{1\,i_{\sigma(m_1)}}
                \cdots \psi^{\dagger i_1}_{1\,i_{\sigma(1)}} 
                X^{\dagger\, i_{m_1+m_2+1}}_{\sigma (m_1+m_2+1)} \cdots \cr
\eea
As far as the bosons go, $X$s occupy slots $m_1+m_2+1$ to $m_1+m_2+n_2$, $Y$s occupy slots $m_1+m_2+n_2+1$ to $m_1+m_2+n_2+n_3$, while
$Z$s occupy slots $m_1+m_2+n_2+n_3+1$ to $m_1+m_2+n_2+n_3+n_1$. The boson slots are not reordered by the $\dagger$.
A straight forward computation now shows that
\bea
\langle \chi_{R,(\vec{r},\vec{s})\vec{\alpha}\vec{\beta}}(Z,Y,X,\psi_1,\psi_2)
        \chi^\dagger_{T,(\vec{t},\vec{u})\vec{\gamma}\vec{\delta}}(Z,Y,X,\psi_1,\psi_2)\rangle\cr
\cr
= \delta_{R T}\prod_{i=1}^3\delta_{r_i t_i}\prod_{j=1}^2 \delta_{s_j u_j}\prod_{k=1}^4 \delta_{\alpha_k\gamma_k}
\prod_{l=1}^4 \delta_{\beta_l \delta_l}
{f_R {\rm hooks}_{R}\over \prod_m {\rm hooks}_{r_m}\prod_n {\rm hooks}_{s_n}}
\label{su23twopoint}
\eea

\section{Action of Dilatation operator in su$(2|3)$ sector}

To simplify the formula for the one loop dilatation operator, set $\phi_1\equiv Z$, $\phi_2\equiv X$ and $\phi_3\equiv Y$.
From the formula (2.1) of \cite{tseytlin}, or the $H_2$ piece of Table 1 of \cite{beisert}, we find the following
one loop dilatation operator 
\bea
  D=&-&g_{YM}^2 \left(\sum_{i>j=1}^3 \, {\rm Tr}\left(\left[\phi_i,\phi_j\right]\left[\partial_{\phi_i},\partial_{\phi_j}\right]\right)
     +\sum_{i=1}^3\sum_{a=1}^2 \, {\rm Tr}\left(\left[\phi_i,\psi_a\right]\left[\partial_{\phi_i},\partial_{\psi_a}\right]\right)\right.
\cr
&+&\,{\rm Tr}\left(\left\{\psi_1,\psi_2\right\}\left\{\partial_{\psi_1},\partial_{\psi_2}\right\}\right)
\Bigg)
\label{fullD}
\eea

We will study the limit in which the number of $\phi_1$s (=$n_1$) is much greater than the number of 
$\phi_2$s (=$n_2$), $\phi_3$s(=$n_3$), $\psi_1$s (=$m_1$) and
$\psi_2$s (=$m_2$). In this limit we can simplify the dilatation operator to
\bea
  D=-g_{YM}^2\left(\sum_{j=2}^3 \, {\rm Tr}\left(\left[\phi_1,\phi_j\right]\left[\partial_{\phi_1},\partial_{\phi_j}\right]\right)
     + \sum_{a=1}^2 \, {\rm Tr}\left(\left[\phi_1,\psi_a\right]\left[\partial_{\phi_1},\partial_{\psi_a}\right]\right)\right)
\label{simpleD}
\eea
The simpler expression (\ref{simpleD}) is obtained from (\ref{fullD}) simply 
by noting that a derivative with respect to $\phi_1$ will generate $n_1$ terms. 
Since $n_1\gg n_2,n_3,m_1,m_2$, this is a lot more terms than is generated by differentiating with
respect to any other field. 

The simplest example to start with is when the operator is built using only one fermion $\psi_1$ and one boson $\phi_1\equiv Z$. 
One of the terms we need to evaluate is
\bea
   Z^i_j\psi^j_{1\, k}{d\over dZ^l_k}{d\over d\psi^i_{1\, l}}\left(
   {1\over n!m!}\sum_{\sigma\in S_{n+m}}{\rm Tr}_{(r,s)\alpha\beta}(\Gamma^R(\sigma))
   \psi^{i_1}_{1\, i_{\sigma(1)}}\cdots \psi^{i_m}_{1\, i_{\sigma(m)}}
   Z^{i_{m+1}}_{i_{\sigma (m+1)}}\cdots Z^{i_{m+n}}_{i_{\sigma (m+n)}}\right)
\eea
To take this derivative we need to use the product rule, and hit each of the $m$ factors of $\psi_1$ and each of the $n$ factors of $Z$. 
We know that the contribution from each $Z$ derivative is the same so that we simply get an overall $n$ multiplied by the term obtained 
when the derivative hits (say) the $Z$ in slot $m+1$. 
The first thing we want to argue is that the contribution from each $\psi_1$ derivative is also the same, so that we can write these
$m$ terms as $m$ multiplied by the term obtained when the derivative hits (say) the $\psi_1$ in slot $1$. To start, think of
\bea
   \psi^j_{1\, k}{d\over d\psi^i_{1\, l}}
   \label{onething}
\eea
as our operator. It is Grassman even so it commutes with all other variables. 
This allows us to move it into any slot, without costing any signs.
Now consider
\bea
   \sum_{\rho\in S_{n+m}}{\rm Tr}\left(P_{R,(r,s)\alpha\beta}\Gamma^R \left( (1,m+1)\rho \right)\right)\delta^{i_1}_{i_{\rho (1)}}
   \psi^{i_{1\, 1}}_{1\, i_{\rho (m+1)}}\psi^{i_2}_{1\, i_{\rho(2)}}\cdots\psi^{i_m}_{1\, i_{\rho(m)}}
   Z^{i_{m+1}}_{i_{\rho (1)}}Z^{i_{m+2}}_{i_{\rho (m+2)}}\cdots Z^{i_{m+n}}_{i_{\rho (m+n)}}\cr
   = \sum_{\rho\in S_{n+m}}{\rm Tr}\left(P_{R,(r,s)\alpha\beta}\Gamma^R \left( (1,m+1)\rho \right)\right)\delta^{i_1}_{i_{\rho (1)}}
   {\rm Tr}_{V^{\otimes\, n+m}}\left(\rho (1,m+1) \psi_1^{\otimes\, m} Z^{\otimes\, n}\right)
   \cr
\eea
We can now change variables from $\rho$ to $\gamma = (1,l)\rho (1,l)$ to obtain
\bea
   =\sum_{\gamma\in S_{n+m}}{\rm Tr}\left(P_{R,(r,s)\alpha\beta}\Gamma^R \left( (1,m+1)(1,l)\gamma (1,l) \right)\right)\delta^{i_l}_{i_{\gamma (l)}}
   {\rm Tr}_{V^{\otimes\, n+m}}\left((1,l)\gamma (1,l)(1,m+1) \psi_1^{\otimes\, m} Z^{\otimes\, n}\right)\cr
\eea
Now,
\bea
{\rm Tr}_{V^{\otimes\, n+m}}\left((1,l)\gamma (1,l)(1,m+1) \psi_1^{\otimes\, m} Z^{\otimes\, n}\right)
={\rm Tr}_{V^{\otimes\, n+m}}\left(\gamma (l,m+1)(1,l) \psi_1^{\otimes\, m} Z^{\otimes\, n}(1,l)\right)\cr
= \psi^{i_l}_{1\, i_{\gamma (m+1)}}\psi^{i_2}_{1\, i_{\gamma(2)}}\cdots 
  \psi^{i_{l-1}}_{1\, i_{\gamma(l-1)}}\psi^{i_1}_{1\, i_{\gamma (1)}}\psi^{i_{l+1}}_{1\,i_{\gamma(l+1)}}\cdots\psi^{i_m}_{1\,i_{\gamma(m)}}
   Z^{i_{m+1}}_{i_{\gamma (l)}}Z^{i_{m+2}}_{i_{\gamma (m+2)}}\cdots Z^{i_{m+n}}_{i_{\gamma (m+n)}}\cr
= -\psi^{i_1}_{1\, i_{\gamma (1)}}\psi^{i_2}_{1\,i_{\gamma(2)}}\cdots 
  \psi^{i_{l-1}}_{1\,i_{\gamma(l-1)}}\psi^{i_l}_{1\, i_{\gamma (m+1)}}
  \psi^{i_{l+1}}_{1\,i_{\gamma(l+1)}}\cdots\psi^{i_m}_{1\, i_{\gamma(m)}}
   Z^{i_{m+1}}_{i_{\gamma (l)}}Z^{i_{m+2}}_{i_{\gamma (m+2)}}\cdots Z^{i_{m+n}}_{i_{\gamma (m+n)}}\cr
\eea
Also,
\bea
   {\rm Tr}\left(P_{R,(r,s)\alpha\beta}\Gamma^R \left( (1,m+1)(1,l)\gamma (1,l) \right)\right)
   ={\rm Tr}\left(\Gamma^R \left( (1,l) \right)P_{R,(r,s)\alpha\beta}\Gamma^R \left( (1,l)(l,m+1)\gamma \right)\right)\cr
   =-{\rm Tr}\left(P_{R,(r,s)\alpha\beta}\Gamma^R \left((l,m+1)\gamma \right)\right)  \cr
\eea
Thus, we find
\bea
&&  \sum_{\rho\in S_{n+m}}{\rm Tr}\left(P_{R,(r,s)\alpha\beta}\Gamma^R \left( (1,m+1)\rho \right)\right)\delta^{i_1}_{i_{\rho (1)}}
   \psi^{i_{1\, 1}}_{1\, i_{\rho (m+1)}}\psi^{i_2}_{1\, i_{\rho(2)}}\cdots\psi^{i_m}_{1\, i_{\rho(m)}}
   Z^{i_{m+1}}_{i_{\rho (1)}}Z^{i_{m+2}}_{i_{\rho (m+2)}}\cdots Z^{i_{m+n}}_{i_{\rho (m+n)}}\cr
&=&\sum_{\gamma\in S_{n+m}}{\rm Tr}\left(P_{R,(r,s)\alpha\beta}\Gamma^R \left((l,m+1)\gamma \right)\right)\delta^{i_l}_{i_{\gamma (l)}}
\psi^{i_1}_{1\, i_{\gamma (1)}}\psi^{i_2}_{1\,i_{\gamma(2)}}\cdots 
  \psi^{i_{l-1}}_{1\,i_{\gamma(l-1)}}\psi^{i_l}_{1\, i_{\gamma (m+1)}}
  \psi^{i_{l+1}}_{1\,i_{\gamma(l+1)}}\cdots\psi^{i_m}_{1\, i_{\gamma(m)}}\cr
 && \times Z^{i_{m+1}}_{i_{\gamma (l)}}Z^{i_{m+2}}_{i_{\gamma (m+2)}}\cdots Z^{i_{m+n}}_{i_{\gamma (m+n)}}
\eea
The LHS of this last identity is obtained when we differentiate the $\psi_1$ in slot 1; the RHS when we differentiate $\psi_1$ in slot $l$.
Thus, this last identity proves that the contribution from each $\psi_1$ derivative is the same. Thus,
\bea
&& Z^i_j\psi^j_{1\, k}{d\over dZ^l_k}{d\over d\psi^i_{1\, l}}\left(
   {1\over n!m!}\sum_{\sigma\in S_{n+m}}{\rm Tr}_{(r,s)\alpha\beta}(\Gamma^R(\sigma))
   \psi^{i_1}_{1\, i_{\sigma(1)}}\cdots \psi^{i_m}_{1\, i_{\sigma(m)}}
   Z^{i_{m+1}}_{i_{\sigma (m+1)}}\cdots Z^{i_{m+n}}_{i_{\sigma (m+n)}}\right)\cr
&=&{1\over (n-1)!(m-1)!}\sum_{\rho\in S_{n+m}}{\rm Tr}\left(P_{R,(r,s)\alpha\beta}
     \Gamma^R \left( (1,m+1)\rho \right)\right)\delta^{i_{1}}_{i_{\rho (1)}}
     \psi^{j}_{1\, i_{\rho (m+1)}}\psi^{i_2}_{1\, i_{\rho(2)}}\cdots\psi^{i_m}_{1\, i_{\rho(m)}}\cr
&\times& Z^{i_{1}}_{j}Z^{i_{m+2}}_{i_{\rho (m+2)}}\cdots Z^{i_{m+n}}_{i_{\rho (m+n)}}
\eea
It is now a simple exercise to find
\bea
  D\chi_{R,(r,s)\alpha\beta}(\psi_1,Z)
=-{2g_{YM}^2\over (4\pi)^2}{\rm Tr}\left(\left[Z,\psi_1\right]\left[\partial_{Z},\partial_{\psi_1}\right]\right)
\chi_{R,(r,s)\alpha\beta}(\psi_1,Z)\cr
={2g_{YM}^2\over (4\pi)^2 (n-1)!(m-1)!} \sum_{\rho\in S_{n+m}}\delta^{i_{1}}_{i_{\rho (1)}}
{\rm Tr}_{(r,s)\alpha\beta}\left(\Gamma^R ([(1,m+1) ,\rho])\right)\times\cr
{\rm Tr}_{V^{\otimes\, n+m}} \left( [(1,m+1),\rho] \psi_1^{\otimes \, m}Z^{\otimes\, n}\right)
\eea
Our next task is to express ${\rm Tr}_{V^{\otimes\, n+m}} \left( [(1,m+1),\rho] \psi_1^{\otimes \, m}Z^{\otimes\, n}\right)$
as a sum over restricted Schur polynomials. 
We will generalize the argument given in \cite{rajmike} which provides the identity for restricted
Schur polynomials built entirely out of bosonic fields. 
First we need an identity. 
The irrep $(r,s)$ of $S_n\times S_m$ will, in general, be subduced by irrep $R$ of $S_{n+m}$ more than once. 
Label these different copies with $\beta$. 
It is convenient to switch to a bra-ket notation. 
In this notation the operators used to define the restricted Schur polynomials we have constructed are
\bea
[P_{R,(r,s)\alpha\beta}]_{JI}=
\sum_{a,b,i}\langle R,J|s,b;r,i;\alpha\rangle\langle s^T,a;r,i;\beta|R,I\rangle O_{ba}
\eea
We will make use of the identity\cite{BHR1,rajmike}
\bea
\sum_{\beta}\langle R,I|r,b;s,i;\beta\rangle\langle r,a;s,j;\beta |R,J\rangle
={d_r d_s\over n!m!}\sum_{\alpha_1\in S_m}\sum_{\alpha_2\in S_n}\Gamma^s (\alpha_1^{-1})_{ij}\Gamma^r (\alpha_2^{-1})_{ab}
\Gamma^R (\alpha_1\circ\alpha_2)_{IJ}\cr
\label{useful}
\eea
in what follows. Consider the sum
\bea
\sum_{R,(r,s)\alpha\beta} {d_R n! m!\over d_r d_s (n+m)!} \chi_{R,(r,s)\alpha\beta} (\tau)\chi_{R,(r,s)\beta\alpha}^\dagger (\sigma)\cr
=\sum_{R,(r,s)\alpha\beta} {d_R n! m!\over d_r d_s (n+m)!}
{\rm Tr}(P_{R,(r,s)\alpha\beta}\Gamma^R(\tau)){\rm Tr}(P^\dagger_{R,(r,s)\beta\alpha}\Gamma^R(\sigma))\cr
=\sum_{R,(r,s)\alpha\beta} {d_R n! m!\over d_r d_s (n+m)!}
[P_{R,(r,s)\alpha\beta}]_{IJ}[\Gamma^R(\tau)]_{JI}[P^\dagger_{R,(r,s)\beta\alpha}]_{KL}[\Gamma^R(\sigma)]_{LK}
\eea
Now, rewrite both projectors using bra-ket notation to find
\bea
\sum_{R,(r,s)\alpha\beta} \sum_{IJ} \sum_{KL}
{d_R n! m!\over d_r d_s (n+m)!} \chi_{R,(r,s)\alpha\beta} (\tau)\chi_{R,(r,s)\beta\alpha}^\dagger (\sigma)\cr
=\sum_{R,(r,s)\alpha\beta} {d_R n! m!\over d_r d_s (n+m)!}
\sum_{a,b,i,I,J} \langle R,I|s,b;r,i;\alpha\rangle\langle s^T,a;r,i;\beta |R,J\rangle O_{ba}
[\Gamma^R(\tau)]_{JI}\cr
\times \sum_{c,d,j,K,L} \langle R,L|s^T,d;r,j;\beta\rangle\langle s,c;r,j;\alpha |R,K\rangle (O^T)_{dc}
[\Gamma^R(\sigma)]_{LK}
\eea
The sum over the multiplicity labels can now be performed using (\ref{useful}). We find
\bea
\sum_{R,(r,s)\alpha\beta} {d_R n! m!\over d_r d_s (n+m)!} \chi_{R,(r,s)\alpha\beta} (\tau)\chi^\dagger_{R,(r,s)\beta\alpha} (\sigma)\cr
= \sum_{R,(r,s)\alpha\beta}\sum_{\gamma_1,\tau_1\in S_m}\sum_{\gamma_2,\tau_2\in S_n}
{d_R d_r d_s\over (n+m)!n!m!}
{\rm Tr}(\Gamma^r (\gamma_2\tau_2)) {\rm Tr}(\Gamma^{s^T}(\gamma_1) O^T\Gamma^s (\tau_1) O)\cr
\times {\rm Tr} (\Gamma^R(\tau\cdot\gamma_1\circ\gamma_2 \cdot\sigma\cdot\tau_1\circ\tau_2) )
\eea
In this last expression we recognize the delta function on the group
\bea
\sum_R {d_R\over |{\cal G}|}\chi_R (\sigma)=\delta (\sigma)
\eea
where $R$ is a complete set of irreps of group ${\cal G}$. Thus, we now have
\bea
\sum_{R,(r,s)\alpha\beta} {d_R n! m!\over d_r d_s (n+m)!} \chi_{R,(r,s)\alpha\beta} (\tau)\chi^\dagger_{R,(r,s)\beta\alpha} (\sigma)\cr
=\sum_{R\vdash n+m}\sum_{\tau_1\in S_m}\sum_{\tau_2\in S_n}{\rm sgn}(\tau_1){d_R\over (n+m)!}
\chi_R (\tau\cdot\tau_1^{-1}\circ\tau_2^{-1} \cdot\sigma\cdot\tau_1\circ\tau_2)\cr
=\sum_{\tau_1\in S_m}\sum_{\tau_2\in S_n}{\rm sgn}(\tau_1)
\delta (\tau\cdot\tau_1^{-1}\circ\tau_2^{-1} \cdot\sigma\cdot\tau_1\circ\tau_2)
\eea
This identity is all that is needed to prove that
\bea
{\rm Tr}_{V^{\otimes\, n+m}}(\sigma \psi^{\otimes\, m}_1 Z^{\otimes \, n})
=\sum_{R,(r,s)\alpha\beta} {d_R n! m!\over d_r d_s (n+m)!} \chi^\dagger_{R,(r,s)\alpha\beta} (\sigma)\chi_{R,(r,s)\beta\alpha} (\psi_1,Z)
\label{nicenewone}
\eea
It is now straight forwards to see that
\bea
D\chi_{R,(r,s)\alpha\beta}(\psi_1,Z)=\sum_{T,(t,u)\gamma\delta} M_{R,(r,s)\alpha\beta;T,(t,u)\gamma\delta}\chi_{T,(t,u)\gamma\delta}(\psi_1,Z)
\eea
where
\bea
M_{R,(r,s)\alpha\beta ;T,(t,u)\gamma\delta} = - g_{YM}^2\sum_{R'}{c_{RR'} d_T n m\over d_{R'} d_t d_u (n+m)}
{\rm Tr}\Big( \Big[ \Gamma^R((1,m+1)),P_{R,(r,s)\alpha\beta}\Big]I_{R'\, T'}\times \cr
\times \Big[\Gamma^T((1,m+1)),P_{T,(t,u)\delta\gamma}\Big] I_{T'\, R'}\Big)\cr
\eea
To obtain the spectrum of anomalous dimensions, it is convenient to consider the action of the dilatation operator on 
operators with two point functions normalized to 1. 
The normalized operators can be obtained from
\bea
  \chi_{R,(r,s)\alpha\beta}(\psi_1,Z)=\sqrt{f_R \, {\rm hooks}_R\over {\rm hooks}_r\, {\rm hooks}_s}O_{R,(r,s)\alpha\beta}(\psi_1,Z)
\eea
In terms of these normalized operators
\bea
DO_{R,(r,s)\alpha\beta}(\psi_1,Z)=\sum_{T,(t,u)\gamma\delta} N_{R,(r,s)\alpha\beta;T,(t,u)\gamma\delta}O_{T,(t,u)\gamma\delta}(\psi_1,Z)
\eea
where
\bea
N_{R,(r,s)\alpha\beta;T,(t,u)\gamma\delta}= - g_{YM}^2\sum_{R'}{c_{RR'} d_T n m\over d_{R'} d_t d_u (n+m)}
\sqrt{f_T \, {\rm hooks}_T\, {\rm hooks}_r \, {\rm hooks}_s \over f_R \, {\rm hooks}_R\, {\rm hooks}_t\, {\rm hooks}_u}\times\cr
\times{\rm Tr}_{R\oplus T}\Big(\Big[ \Gamma^R((1,m+1)),P_{R,(r,s)\alpha\beta}\Big]I_{R'\, T'}
\Big[\Gamma^T((1,m+1)),P_{T,(t,u)\delta\gamma}\Big]I_{T'\, R'}\Big) \cr
\label{dilonezonepsi}
\eea
We have explicitly indicated that the last trace is taken over the direct sum of the carrier spaces of $R$ and $T$.
Remarkably, this takes a very similar form as the action of the dilatation operator in the $SU(2)$ sector\cite{DeComarmond:2010ie}.
Consequently, we know that the operators with a definite scaling dimension can be constructed using the ideas of the double coset 
ansatz of \cite{DCI}. A few of the details are different, so it is worth describing some of the steps involved. We will however
freely draw on \cite{DCI}, so the reader wanting to follow the details is encouraged to study \cite{DCI} in detail.

As already described above, we remove boxes from $R$ to produce $r$. 
$m_i$ is the number of boxes that must be removed from row $i$ of $R$ to produce $r$.
The $m_i$ can be assembled to produce the vector label $\vec{m}$ which, as will become clear, is conserved
by the one loop dilatation operator. The subgroup
\bea
  H=S_{m_1}\times S_{m_2}\times \cdots \times S_{m_p}
\eea
will play an important role is what follows.
We make use of two types of branching coefficients
\bea
  \sum_{\mu}B^{s\to 1_H}_{k\mu}B^{s\to 1_H}_{l\mu} = {1\over |H|}\sum_{\gamma\in H}\Gamma^{s}(\gamma)_{kl}
\eea
\bea
  \sum_{\mu}B^{s^T\to 1^m}_{k\mu}B^{s^T\to 1^m}_{l\mu} = {1\over |H|}\sum_{\gamma\in H}{\rm sgn}(\gamma)\Gamma^{s^T}(\gamma)_{kl}
\eea
The branching coefficients $B^{s\to 1_H}_{l\mu}$ resolve the multiplicities that arise when we restrict irrep $s$ of $S_m$
to the identity representation $1_H$ of $H$ for which $\Gamma^{1_H}(\gamma)=1$ $\forall\gamma$.  
The branching coefficients $B^{s\to 1^m}_{l\mu}$ resolve the multiplicities that arise when we restrict irrep $s$ of $S_m$
to the representation $1^m$ of $H$ for which $\Gamma^{1^m} (\gamma)={\rm sgn}(\gamma)$ $\forall\gamma$.  
Notice that
\bea
\sum_{\mu}B^{s^T\to 1^m}_{k\mu}B^{s^T\to 1^m}_{l\mu} 
&=& {1\over |H|}\sum_{\gamma\in H}{\rm sgn}(\gamma)\Gamma^{s^T}(\gamma)_{kl}\cr
&=& {1\over |H|}\sum_{\gamma\in H} \left(O^T\Gamma^{s}(\gamma)O\right)_{kl}\cr
&=& O^T_{km} \sum_{\mu}B^{s\to 1_H}_{m\mu}B^{s\to 1_H}_{n\mu}O_{nl}
\eea
so that we can identify $B^{s\to 1_H}_{n\mu}O_{nl}=B^{s^T\to 1^m}_{l\mu}$.
This argument suggests that the multiplicity problem of $s\to 1_H$ can be identified with the
multiplicity problem of $s^T\to 1^m$. 
It is simple to prove that this is indeed the case as follows: Denote the multiplicity of
$1_H$ in $s$ by $n^s_{1_H}$ and the multiplicity of $1^m$ in $s^T$ by $n^{s^T}_{1^m}$. We have
\bea
  n^s_{1_H}&=&{1\over |H|}\sum\chi_s(\sigma)\chi_{1_H}(\sigma)\cr
           &=&{1\over |H|}\sum\chi_s(\sigma)\cr
           &=&{1\over |H|}\sum\chi_{s^T}(\sigma){\rm sgn}(\sigma)\cr
           &=&{1\over |H|}\sum\chi_{s^T}(\sigma)\chi_{1^m}(\sigma)\cr
           &=&n^{s^T}_{1^m}
\eea
which completes the demonstration. Now, following\cite{DCI} we identify
\bea
  |\vec{m},s,\mu;i\rangle =\sum_j B^{s\to 1_H}_{j\mu}\sum_{\sigma\in S_m}\Gamma^s(\sigma)_{ij}|v_\sigma\rangle
\eea
The components $m_i$ of the vector label $\vec{m}$ appearing in the above ket record the number of boxes that
must be removed from row $i$ of $R$ to produce $r$.
These are the basis vectors in $s$ that are used to construct the projectors appearing in the restricted Schur polynomials. 
To construct the projectors, we also need to make use of a basis for $s^T$. 
The basis for $s^T$ should be constructed using $\hat{O}^T$ which provides a map from the carrier space of $s^T$ to the carrier space of $s$. 
Using $\hat{O}^T$ we find
\bea
  \sum_i (\hat{O}^T)_{ki}|\vec{m},s,\mu;i\rangle =\sum_j B^{s^T\to 1^m}_{j\mu}\sum_{\sigma\in S_m}{\rm sgn}(\sigma)\Gamma^{s^T}(\sigma)_{ij}
                                                        |v_\sigma\rangle
\eea
Given these bases, it is a simple matter to verify that the projectors appearing in the restricted Schur polynomials can be written as
\bea
O(s\alpha,s^T\beta) = {d_s \over m! |H|}\sum_{\sigma\, ,\, \tau\in S_m}
B^{s\to 1_H}_{c\alpha}\Gamma^{s}_{ac}(\sigma) |v_\sigma\rangle
\langle v_\tau|B^{s^T\to 1^m}_{d\beta}{\rm sgn}(\tau)\Gamma^{s^T}_{bd}(\tau)O^T_{ba}
\eea
Indeed, using these expressions it is straight forwards to verify that
\bea
   O(s\alpha,s^T\beta) O(s^T\beta,s\alpha) = {\bf 1}_{s}
\qquad
   O(s^T\beta,s\alpha) O(s\alpha,s^T\beta) = {\bf 1}_{s^T}
\eea

In terms of the branching coefficients, introduce the quantities
\bea
   C^{(s)}_{\mu_1\mu_2}(\tau)=|H|\sqrt{d_s\over m!}\left(\Gamma^{s}(\tau)\hat{O}\right)_{km}
                                 B^{s\to 1_H}_{k\mu_1}B^{s^T\to 1^m}_{m\mu_2}
\eea
It is simple to verify that these coefficients define an orthogonal transformation
\bea
C^{(s)}_{\mu_1\mu_2}(\tau)C^{(s)}_{\mu_1\mu_2}(\sigma)&=&\sum_{s\vdash m}\sum_{\gamma_1,\gamma_2\in H}
|H|^2{d_s\over m!}{\rm sgn}(\gamma_2){\rm Tr}(\Gamma^{s}(\tau)\hat{O}\Gamma^{s^T}(\gamma_2)\hat{O}^T\Gamma^{s}(\sigma^{-1})
\Gamma^{s}(\gamma_1))\cr
&=&\sum_{s\vdash m}\sum_{\gamma_1\,\gamma_2\in H}{d_s\over m!}\chi_s (\tau \gamma_2 \sigma^{-1} \gamma_1)\cr
&=&\sum_{\gamma_1\,\gamma_2\in H}\delta (\tau \gamma_2 \sigma^{-1} \gamma_1)
\label{orthogonality}
\eea
It is then rather natural to build operators dual to Gauss configuration $\sigma$ by
\bea
  O_{R,r}(\sigma)=\sum_{s\vdash m}\sum_{\mu_1,\mu_2}C^{(s)}_{\mu_1\mu_2}(\sigma) O_{R,(r,s)\mu_1\mu_2}
  \label{gaussgraphoperators}
\eea
Using (\ref{orthogonality}) we easily find
\bea
  \langle O_{R,r}(\sigma_1)O_{T,t}^\dagger (\sigma_2) \rangle =\sum_{\gamma_1,\gamma_2\in H}\delta (\gamma_1\sigma_1\gamma_2\sigma_2^{-1})
\eea

We ultimately want to evaluate the action of the dilatation operator on the Gauss graph operators (\ref{gaussgraphoperators}).
To do this we need to review the evaluation of the dilatation operator on normalized restricted Shur
polynomials $O_{R,(r,s)\mu_1\mu_2}$. For all the details see \cite{Koch:2011hb}. 
Denote the number of rows in the Young diagram labeling the restricted Schur polynomial by $p$.
The one loop dilatation operator given in (\ref{dilonezonepsi}) is exact to all order in $1/N$.
To capture the large $N$ (but not planar!) limit we use the displaced corners approximation.
Recall that to subduce $r\vdash n$ from $R\vdash m+n$ we remove $m$ boxes from $R$.
Each box in the Young diagram $R$ can be assigned a factor which is equal to $N-i+j$ for the box in row $i$ and column $j$.
The displaced corners approximation applies when the difference between the factors of any two boxes (of the $m$ boxes removed)
is of order $N$ whenever the removed boxes come from different rows.
The action of the dilatation operator simplifies in this limit because the action of the symmetric group becomes particularly simple\cite{mn}.
When the displaced corners approximation holds, we associate each removed box with a vector in a $p$ dimensional vector space $V_p$.
In this way the $m$ removed boxes associated with the $\psi_1$'s define a vector in $V_p^{\otimes\, m}$.
The trace over $R\oplus T$ factorizes into a trace over $r\oplus t$ and a trace over $V_p^{\otimes m}$. 
The bulk of the work is in evaluating the trace over $V_p^{\otimes m}$. 
This trace is evaluated using the methods developed in \cite{Koch:2011hb} as we now explain.
Introduce a basis for the fundamental representation of the Lie algebra u$(p)$ given by $(E_{ij})_{ab}=\delta_{ia}\delta_{jb}$.
These Lie algebra elements obey
\bea
  E_{ij}E_{kl} = \delta_{jk} E_{il}
\eea
If a box is removed from row $i$ it is associated to a vector $v_i$ which is an eigenstate of $E_{ii}$ with eigenvalue 1.
The intertwining maps can be written in terms of the $E_{ij}$. For example, if we remove a box from
row $i$ of $R$ and a box from row $j$ of $T$, assuming that $R'$ and $T'$ have the same shape, we have
\bea
  I_{T'R'}=E^{(1)}_{ji}
\eea
By realizing the intertwiners in this way, we find a simple result for the product of symmetric group elements with the intertwiners. 
For example,
\bea
  (1,m+1)I_{T'R'} \, = \, E^{(1)}_{kl}E^{(m+1)}_{lk}E^{(1)}_{ji} \, = \, E^{(1)}_{ki}E^{(m+1)}_{jk}
\eea
Using these techniques we find\cite{Koch:2011hb}
\bea
  D O_{R,(r,s)\mu_1\mu_2}=-g_{YM}^2\sum_{u\nu_1\nu_2}\sum_{i<j}\delta_{\vec{m},\vec{n}}M^{(ij)}_{s\mu_1\mu_2 ; u\nu_1\nu_2}\Delta_{ij}
                           O_{R,(r,u)\nu_1\nu_2}
  \label{factoredD}
\eea
where $\Delta_{ij}$ acts only on the Young diagrams $R,r$ and
\bea
  M^{(ij)}_{s\mu_1\mu_2 ; u\nu_1\nu_2}&&={m\over\sqrt{d_s d_u}}\left({\rm Tr}
       \left(\hat{O}(s\mu_1;s^T\mu_2) E^{(1)}_{ii} \hat{O}(u^T\nu_2;u\nu_1)E^{(1)}_{jj} \right)\right.\cr
&&\left. +
       {\rm Tr}\left(\hat{O}(s\mu_1;s^T\mu_2) E^{(1)}_{jj} \hat{O}(u^T\nu_2;u\nu_1)E^{(1)}_{ii} \right)\right)
\eea
The operator $\Delta_{ij}$ splits into three terms
\bea
  \Delta_{ij}=\Delta_{ij}^{+}+\Delta_{ij}^{0}+\Delta_{ij}^{-}
\eea
To describe the action of these three pieces, we will need a little more notation.
Denote the row lengths of $r$ by $r_i$. 
The Young diagram $r_{ij}^+$ is obtained by removing a box from row $j$ and adding it to row $i$
and $r_{ij}^-$ is obtained by removing a box from row $i$ and adding it to row $j$.
We now have
\bea
  \Delta_{ij}^{0}O_{R,(r,s)\mu_1\mu_2} = -(2N+r_i+r_j)O_{R,(r,s)\mu_1\mu_2}
  \label{0term}
\eea
\bea
  \Delta_{ij}^{+}O_{R,(r,s)\mu_1\mu_2} = \sqrt{(N+r_i)(N+r_j)}O_{R^+_{ij},(r^+_{ij},s)\mu_1\mu_2}
  \label{pterm}
\eea
\bea
  \Delta_{ij}^{-}O_{R,(r,s)\mu_1\mu_2} = \sqrt{(N+r_i)(N+r_j)}O_{R^-_{ij},(r^-_{ij},s)\mu_1\mu_2}
  \label{mterm}
\eea
It is clear that the dilatation operator factorized into a piece ($\Delta_{ij}$) that acts only on $r$ (i.e. on $Z$s) and 
a piece $M^{(ij)}_{s\mu_1\mu_2 ; u\nu_1\nu_2}$ that acts only on $s$ (the $\psi_1$s).
Further, because $R$ and $r$ change in exactly the same way the vector $\vec{m}$ is preserved by the dilatation operator. 
We are now ready to consider the action of the dilatation operator on the Gauss graph operators (\ref{gaussgraphoperators}).
Towards this end, consider
\bea
\langle O^\dagger_{T,t}(\sigma_2)DO_{R,r}(\sigma_1)\rangle
&=& {|H|^2\over m!}\sum_{s,u\vdash m}\sum_{\mu_1 \mu_2 \nu_1 \nu_2}\sqrt{d_s d_u}
\left(\Gamma^s(\sigma_1)\hat{O}\right)_{k_1 m_1}B^{s\to 1_H}_{k_1\mu_1}B^{s^T\to 1^m}_{m_1 \mu_2}\cr
&&\quad\times \left(\Gamma^u (\sigma_2)\hat{O}\right)_{k_2 m_2}B^{u\to 1_H}_{k_2\nu_1}B^{u^T\to 1^m}_{m_2\nu_2}
\langle O^\dagger_{T,(t,u)\nu_1\nu_2}\, D\, O_{R,(r,s)\mu_1\mu_2}\rangle\cr
&=&-{|H|^2\over m!}g_{YM}^2\sum_{s,u\vdash m}\sum_{\mu_1 \mu_2 \nu_1 \nu_2}
\left(\Gamma^s(\sigma_1)\hat{O}\right)_{k_1 m_1}B^{s\to 1_H}_{k_1\mu_1}B^{s^T\to 1^m}_{m_1 \mu_2}\cr
&&\quad\times \left(\Gamma^u (\sigma_2)\hat{O}\right)_{k_2 m_2}B^{u\to 1_H}_{k_2\nu_1}B^{u^T\to 1^m}_{m_2\nu_2}
\sum_{i<j}\Delta_{ij}^{R,r;T,t}\, m\times\cr
&&\times\left(
\langle\vec{m},s^T,\mu_2;a|E_{ii}^{(1)}|\vec{m},u^T,\nu_2;b\rangle
\langle\vec{m},u,\nu_1;b|E_{jj}^{(1)}|\vec{m},s,\mu_1;a\rangle +\right.\cr
&&\left.\langle\vec{m},s^T,\mu_2;a|E_{jj}^{(1)}|\vec{m},u^T,\nu_2;b\rangle
\langle\vec{m},u,\nu_1;b|E_{ii}^{(1)}|\vec{m},s,\mu_1;a\rangle
\right)
\eea
Now, focus on the evaluation of
\bea
&&\sum_u\sum_{\nu_1,\nu_2} |\vec{m},u^T,\nu_2;b\rangle \langle\vec{m},u,\nu_1;b|
\left(\Gamma^u (\sigma_2)\hat{O}\right)_{k_2 m_2}B^{u\to 1_H}_{k_2\nu_1}B^{u^T\to 1^m}_{m_2\nu_2}\cr
&&\quad
=\sum_u\sum_{\nu_1,\nu_2}\sum_{\sigma,\tau\in S_m} {d_u\over |H|m!}
{\rm sgn}(\tau)B_{d\nu_2}^{u^T\to 1^m}\Gamma^{u^T}_{bd}(\tau)|v_\tau\rangle O_{cb}
\langle v_\sigma |\Gamma^u_{ce}(\sigma)B^{u\to 1_H}_{e\nu_1}\cr
&&\qquad\times \left(\Gamma^u (\sigma_2)\hat{O}\right)_{k_2 m_2}B^{u\to 1_H}_{k_2\nu_1}B^{u^T\to 1^m}_{m_2\nu_2}\cr
&=& \sum_u\sum_{\sigma,\tau\in S_m}\sum_{\gamma_1,\gamma_2\in H}
 {d_u\over |H|^3 m!} {\rm sgn}(\tau) {\rm sgn}(\gamma_2)|v_\tau\rangle\langle v_\sigma |
{\rm Tr}(\Gamma^u(\gamma_1)\Gamma^u(\sigma^{-1})\hat{O}\Gamma^{u^T}(\tau)\Gamma^{u^T}(\gamma_2)\hat{O}^T\Gamma^u(\sigma_2^{-1}))\cr
&=& \sum_{\gamma_1,\gamma_2\in H}\sum_u\sum_{\sigma,\tau\in S_m} {d_u\over |H|^3 m!} |v_\tau\rangle\langle v_\sigma |
\chi_u(\gamma_1\sigma^{-1}\tau\gamma_2\sigma_2^{-1})\cr
&=& \sum_{\gamma_1,\gamma_2\in H}\sum_{\sigma,\tau\in S_m} {1\over |H|^3} |v_\tau\rangle\langle v_\sigma |
\delta(\gamma_1\sigma^{-1}\tau\gamma_2\sigma_2^{-1})
\eea
From this point on the evaluation proceeds exactly as in \cite{DCI}. The result is
\bea
\langle O_{T,s}^\dagger (\sigma_2)DO_{R,r}(\sigma_1) \rangle =
   - g_{YM}^2\sum_{ \gamma_1, \gamma_2 } 
\delta ( \gamma_1 \sigma_2 \gamma_2  \sigma_1^{-1}  )\sum_{i<j}   ~~ n_{ij} (  \sigma_1  )
  \Delta^{R,r ;  T,s}_{ij}  
 \label{bigcomputed}
\eea
or, equivalently
\bea
D O_{R,r}(\sigma_1)  =
   - g_{YM}^2 \sum_{i<j}   ~~ n_{ij} (  \sigma_1  )
 \Delta_{ij}  O_{R,r}(\sigma_1)
 \label{lovelyanswer}
\eea
This proves that the operators (\ref{gaussgraphoperators}) do indeed diagonalize the impurity labels.
The remaining eigenproblem that must be solved has been studied in detail in \cite{gs}. From the results
of \cite{gs} we know that the spectrum of $D$ reduces to the spectrum of a set of decoupled oscillators,
signaling integrability.

Now consider the general case with multiple bosons $\phi_1,\phi_2,\phi_3$ and fermions $\psi_1,\psi_2$.
Using the methods and results that have already been established, it is straight forward to find that
the dilatation operator (\ref{simpleD}) becomes
\bea
D O_{R,(\vec{r},\vec{s})\vec{\alpha}\vec{\beta}}=\sum_{T,(\vec{t},\vec{u})\vec{\gamma}\vec{\delta}}
N_{R,(\vec{r},\vec{s})\vec{\alpha}\vec{\beta};T,(\vec{t},\vec{u})\vec{\gamma}\vec{\delta}}\,
O_{T,(\vec{t},\vec{u})\vec{\gamma}\vec{\delta}}
\eea
where 
$$
   N_{R,(\vec{r},\vec{s})\vec{\alpha}\vec{\beta};T,(\vec{t},\vec{u})\vec{\gamma}\vec{\delta}} =
  -g_{YM}^2\sum_{R'}{c_{RR'} d_T n_1\over d_{R'}\prod_n d_{t_n} \prod_m d_{u_m} (n_1 + K)}
$$
$$
       \times  
\sqrt{f_T {\rm hooks}_T \prod_a{\rm hooks}_{r_a}\prod_b {\rm hooks}_{s_b}\over f_R {\rm hooks}_R \prod_c{\rm hooks}_{t_c}\prod_d {\rm hooks}_{u_d}}
$$
$$
\times\left[
m_1 {\rm Tr}\left(
\left[\Gamma^R (1,K+1),P_{R,(\vec{r},\vec{s})\vec{\alpha}\vec{\beta}}\right]I_{R'T'}
\left[\Gamma^T (1,K+1),P_{T,(\vec{t},\vec{u})\vec{\gamma}\vec{\delta}}\right]I_{T'R'}
\right)\right.
$$
$$
 +
m_2 {\rm Tr}\left(
\left[\Gamma^R (m_1+1,K+1),P_{R,(\vec{r},\vec{s})\vec{\alpha}\vec{\beta}}\right]I_{R'T'}
\left[\Gamma^T (m_1+1,K+1),P_{T,(\vec{t},\vec{u})\vec{\gamma}\vec{\delta}}\right]I_{T'R'}
\right)
$$
$$
 +
n_2 {\rm Tr}\left(
\left[\Gamma^R (m_1+m_2+1,K+1),P_{R,(\vec{r},\vec{s})\vec{\alpha}\vec{\beta}}\right]I_{R'T'}
\left[\Gamma^T (m_1+m_2+1,K+1),P_{T,(\vec{t},\vec{u})\vec{\gamma}\vec{\delta}}\right]I_{T'R'}
\right)
$$
$$
 +\left.
n_3 {\rm Tr}\left(
\left[\Gamma^R (K-n_3,K+1),P_{R,(\vec{r},\vec{s})\vec{\alpha}\vec{\beta}}\right]I_{R'T'}
\left[\Gamma^T (K-n_3,K+1),P_{T,(\vec{t},\vec{u})\vec{\gamma}\vec{\delta}}\right]I_{T'R'}
\right)\right]
$$
and $K=n_2 + n_3 + m_1 + m_2$ is the total number of impurities. The projectors
$P_{R,(\vec{r},\vec{s})\vec{\alpha}\vec{\beta}}$ and $P_{T,(\vec{t},\vec{u})\vec{\gamma}\vec{\delta}}$
which appear in the above formula have been defined in (\ref{completeprojector}). Notice that these projectors
factorize into a product of factors and further, that in each term above the product of all but the $Z$ projector
and one other have a trivial action. As an example, consider the trace
\bea
T = {\rm Tr}\left(
\left[\Gamma^R (1,K+1),P_{R,(\vec{r},\vec{s})\vec{\alpha}\vec{\beta}}\right]I_{R'T'}
\left[\Gamma^T (1,K+1),P_{T,(\vec{t},\vec{u})\vec{\gamma}\vec{\delta}}\right]I_{T'R'}
\right)
\eea
The swap $(1,K+1)$ only has a non-trivial action on slots $1$ and $K+1$.
Slot 1 is populated by a $\psi_1$ field and corresponds to representation $s_1$.
Slot $K+1$ is populated by a $\phi_1=Z$ field and corresponds to representation $r_1$.
The traces over $r_2,r_3,s_2$ are trivial while the trace over $r_1\oplus s_1$ is performed
exactly as described above. In the end we find
\bea
T=d_{s_2}d_{r_2}d_{r_3}d_{r_1'}\delta_{s_2u_2}\delta_{r_2 t_2}\delta_{r_3 t_3}\delta_{r_1' t_1'} \left({\rm Tr}
       \left(\hat{O}(s_1\mu_1;s_1^T\mu_2) E^{(1)}_{ii} \hat{O}(u_1^T\nu_2;u_1\nu_1)E^{(1)}_{jj} \right) \right.\cr
\left. +
       {\rm Tr}\left(\hat{O}(s_1\mu_1;s_1^T\mu_2) E^{(1)}_{jj} \hat{O}(u_1^T\nu_2;u_1\nu_1)E^{(1)}_{ii} \right)\right)
\eea
Defining the Gauss graph operators in this general case now involves an element of a double coset for each
impurity type. Denote the total number of $(\phi_2,\phi_3,\psi_1,\psi_2)$ impurities by $(n_2,n_3,m_1,m_2)$
and describe the number of boxes removed from row $i$ of $R$ for each impurity type by the vectors
$(\vec{n}_2,\vec{n}_3,\vec{m}_1,\vec{m}_2)$\footnote{Thus, $\vec{m}_2$ has components $(m_2)_i$ with $i=1,2,...,p$
and $\sum_i (m_2)_i = m_2$.}. By $H_{\vec{n}_2}$ (for example) we mean the following group
\bea
   H_{\vec{n}_2}=S_{(n_2)_1}\times S_{(n_2)_2}\times \cdots\times S_{(n_2)_p}
\eea
The relevant double cosets are
\bea
\phi_2&\leftrightarrow& \sigma_{\phi_2}\in H_{\vec{n}_2}\setminus S_{n_2}/H_{\vec{n}_2}\cr
\phi_3&\leftrightarrow& \sigma_{\phi_3}\in H_{\vec{n}_3}\setminus S_{n_3}/H_{\vec{n}_3}\cr
\psi_1&\leftrightarrow& \sigma_{\psi_1}\in H_{\vec{m}_1}\setminus S_{m_1}/H_{\vec{m}_1}\cr
\psi_2&\leftrightarrow& \sigma_{\psi_2}\in H_{\vec{m}_2}\setminus S_{m_2}/H_{\vec{m}_2}
\eea
The orthogonal transformation from the restricted Schur basis to the Gauss graph basis uses both
the group theoretic coefficients of \cite{DCI}
\bea
  C^{(r_i)}_{\mu_1\mu_2}=|H_{\vec{n}_i}|\sqrt{d_{r_i}\over n_i!}\Gamma^{(r_i)}(\tau)_{km}
   B^{r_i\to 1_{H_{\vec{n}_i}}}_{k\mu_1}B^{r_i\to 1_{H_{\vec{n}_i}}}_{m\mu_2}
\eea
to transform the $\phi_2$ and $\phi_3$ labels, and the group theoretic coefficients we have introduced above
\bea
  C^{(s_i)}_{\mu_1\mu_2}=|H_{\vec{m}_i}|\sqrt{d_{s_i}\over m_i!}\left(\Gamma^{(s_i)}(\tau)\hat{O}\right)_{km}
   B^{s_i\to 1_{H_{\vec{m}_i}}}_{k\mu_1}B^{s_i^T\to 1^{m_i}_{H_{\vec{m}_i}}}_{m\mu_2}
\eea
In terms of these coefficients, the Gauss graph operators are
\bea
  O_{R,r_1}(\vec{\sigma})=\sum_{r_2\vdash n_2}\sum_{r_3\vdash n_3}\sum_{s_1\vdash m_1}\sum_{s_2\vdash m_2}\sum_{\vec{\mu},\vec{\nu}}
C^{(r_2)}_{\mu_1\nu_1}(\sigma_{\phi_2})
C^{(r_3)}_{\mu_2\nu_2}(\sigma_{\phi_3})
C^{(s_1)}_{\mu_3\nu_3}(\sigma_{\psi_1})
C^{(s_2)}_{\mu_4\nu_4}(\sigma_{\psi_2})
O_{R,(\vec{r},\vec{s})\vec{\mu}\vec{\nu}}\cr
\eea
The dilatation operator in the Gauss graph basis is
\bea
D O_{R,r_1}(\sigma)  =   - g_{YM}^2 \sum_{i<j}   ~~ 
 \left( n_{ij}(\sigma_{\phi_2})+n_{ij}(\sigma_{\phi_3})+n_{ij}(\sigma_{\psi_1})+n_{ij}(\sigma_{\psi_2})\right)
 \Delta_{ij}  O_{R,r_1}(\sigma_1)
 \label{completelovelyanswer}
\eea
Clearly the result of \cite{gs} again imply that the spectrum of $D$ reduces to a set of decoupled oscillators.
This is a clear indication of integrability in this large $N$ limit of the su$(2|3)$ sector.

\section{Discussion}

We have studied the large $N$ limit of the correlation functions of a class of operators that are AdS/CFT dual to systems 
of excited AdS giant gravitons. 
This large $N$ limit does not coincide with the planar limit and we are forced to sum non-planar contributions.
Our study has included adjoint fermions for the first time. 
To accomplish this, we have explained how to construct restricted Schur polynomials that include both adjoint bosons and adjoint fermions. 
These polynomials diagonalize the free field two point functions to all orders in $N$ and are a complete set of local operators.
We have explored the one loop anomalous dimensions of these operators. 
Our study has proved that the action of the one loop dilatation operator acting
on a sector that includes fermionic fields, is diagonalized by a natural extension of the double coset ansatz of \cite{DCI}. 
The resulting spectrum is identical to the spectrum of a set of decoupled oscillators
which is a clear indication of integrability in this large $N$ limit of the su$(2|3)$ sector.
In an appendix we have also argued that the
double coset ansatz diagonalizes the one loop dilatation operator in the sl$(2)$ sector.
Our results suggest that the double coset ansatz of \cite{DCI} together with the extension described in this
article, may diagonalize the complete one loop dilatation operator.
It would be nice to verify if this is indeed the case.

\noindent
{\it Acknowledgements:}
We would like to thank Sanjaye Ramgoolam for extremely helpful discussions and suggestions.
This work is based upon research supported by the South African Research Chairs
Initiative of the Department of Science and Technology and National Research Foundation.
Any opinion, findings and conclusions or recommendations expressed in this material
are those of the authors and therefore the NRF and DST do not accept any liability
with regard thereto.

\begin{appendix}

\section{Action of Dilatation operator in sl$(2)$ sector}

A restricted Schur polynomial basis for the sl$(2)$ sector was constructed in \cite{deMelloKoch:2011vn}.
The operators are built using  $n$ $Z$'s and $m$ vector ``impurities'', that is, $m$ covariant derivatives $D_+$ act 
on the $n$ $Z$ fields. These operators do not mix with other operators under the action of the 
dilatation operator - they form the closed sl$(2)$ subsector\cite{beisert2}. The impurities are
$Z^{(i)}$, $i=0,1,2,...,m$ where
$$
  Z^{(n)}={1\over n!}D_+^n Z,\qquad Z^{(n)\,\,\dagger} = {1\over n!}D_-^n Z^\dagger
$$
and $Z^{(0)}\equiv Z$. Denote the number of $Z^{(i)}$ by $n_i$.
The restricted Schur polynomial is
\bea
\chi_{R,\{ r_i\}\alpha\beta}(Z^{(0)},Z^{(1)},...,Z^{(m)})=\prod_{k=0}^m {1\over n_k !}\sum_{\sigma\in S_{n_Z}}\chi_{R,\{ r_i\}\alpha\beta}(\sigma )
{\rm Tr}(\sigma \prod_{j=0}^m (Z^{(j)})^{\otimes n_j})\, .
\label{restschur}
\eea
The label $\{ r_i\}\alpha\beta$ specifies an irreducible representation of $S_{n_0}\times S_{n_1}\times\cdots\times S_{n_m}$. It consists of
less than $m$ Young diagrams $\{ r_i\}$ together with a pair of multiplicity labels $\alpha\beta$. 
A given $S_{n_0}\times S_{n_1}\times\cdots\times S_{n_m}$
irreducible representation may be subduced more than once; the multiplicity labels tell us which of the degenerate copies are being
used by the restricted character $\chi_{R,\{ r_i\}}(\sigma )$. The free two point function is
\begin{equation}
\langle \chi_{R,\{ r_i\}\alpha\beta}(P)\chi_{S,\{ s_j\}\delta\gamma}^\dagger (Q)\rangle=\delta_{RS}
\delta_{\{ r_i\}\{ s_j\}} \delta_{\alpha\gamma}\delta_{\beta\delta}{{\rm (hooks)}_R\over {\rm (hooks)}_{\{ r_i\} }}f_R\, .
\label{result}
\end{equation}
The delta function $\delta_{\{ r_i\}\{ s_j\}}$ is 1 if the two $S_{n_0}\times S_{n_1}\times\cdots\times S_{n_m}$
irreducible representations specified by $\{ r_i\}$ and $\{ s_j\}$ are identical; multiplicity labels must also 
match - see \cite{Bhattacharyya:2008rb} for more details. The number ${\rm (hooks)}_{R}$ is the product of the hook lengths for Young diagram 
$R$. The number ${\rm (hooks)}_{\{ r_i\} }$ is the product of the ${\rm (hooks)}_{r_i}$, one factor for each of the $r_i$
appearing in $\{ r_i\}$.
A little work now shows that
$$
  D \chi_{R,(r,s)\alpha\beta }(Z,Z^{(q)})=\sum_{S,(t,u)\gamma\delta}M_{R,(r,s),\alpha\beta\, S,(t,u)\delta\gamma}\chi_{S,(t,u)\delta\gamma}(Z,Z^{(q)})
$$
where
$$
M_{R,(r,s)\alpha\beta ;S,(t,u)\delta\gamma} ={1\over q} M_{R,(r,s)\alpha\beta ;S,(t,u)\delta\gamma}^{\rm su(2)} +
\delta M_{R,(r,s)\alpha\beta ;S,(t,u)\delta\gamma}
$$
$M_{R,(r,s)\alpha\beta ;S,(t,u)\delta\gamma}^{\rm su(2)}$ is identical to the usual action of the dilatation operator in the su$(2)$
sector, while the correction is
$$
 \delta M_{R,(r,s)\alpha\beta\,\, S,(t,u)\delta\gamma}=g_{YM}^2\left({1\over q}-\sum_{i=1}^q {1\over i}\right)\delta_{RS}
\delta_{(r,s)(t,u)}{nm\over d_r d_s}\times
$$
$$
\times\left( \delta_{\alpha\delta}\chi_{R,(r,s)\beta\gamma}\left((1,m+1)\right)
+ \delta_{\beta\gamma}\chi_{R,(r,s)\alpha\delta}\left((1,m+1)\right)\right)
$$
$$
-g_{YM}^2\left({1\over q}-\sum_{i=1}^q {1\over i}\right)\sum_{R'}{c_{RR'}d_S n m \over d_t d_u (n+m) d_{R'}}
\left[
{\rm Tr}(I_{S'R'} P_{R\to (r,s)\alpha\beta}(1,m+1)I_{R'S'}(1,m+1)P_{S\to (t,u)\gamma\delta})\right.
$$
$$ + \left. {\rm Tr}(I_{S'R'} (1,m+1)P_{R\to (r,s)\alpha\beta}I_{R'S'}P_{S\to (t,u)\gamma\delta}(1,m+1))
\right]\, .
$$
Since $M_{R,(r,s)\alpha\beta ;S,(t,u)\delta\gamma}^{\rm su(2)}$ is the usual action of the dilatation operator in the su$(2)$
sector, we know that moving to the Gauss graph basis will diagonalize $M_{R,(r,s)\alpha\beta ;S,(t,u)\delta\gamma}^{\rm su(2)}$
on its impurity labels, leaving only the eigenproblem considered in \cite{gs}. Denoting the piece of the dilatation that leads
to $\delta M_{R,(r,s)\alpha\beta ;S,(t,u)\delta\gamma}$ by $\delta D$ we find that, in the Gauss graph basis
\bea
   \delta D O_{R,r}(\sigma) = 2g_{YM}^2 Nm\left( \sum_{i=1}^q {1\over i}-{1\over q}\right) O_{R,r}(\sigma)
\eea
Thus, the
double coset ansatz diagonalizes the one loop dilatation operator in the sl$(2)$ sector.

\end{appendix}

\end{document}